\input{aipcheck}

\documentclass[
    ,final            
  ]
  {aipproc}

\layoutstyle{6x9}

\begin{document}

\title{Small-${\it x}$ and Forward Measurements at STAR}


\author{Chris Perkins for the STAR Collaboration}{
  address={UC Berkeley/Space Sciences Lab, Stony Brook University}
}

\begin{abstract}
Measurements of azimuthal differences between forward di-pions are sensitive to the low-${\it x}$ gluon content of the proton and provide the best opportunity to probe for gluon saturation in nuclei.
Previously reported analyses have shown that the gluon saturation regime may have been reached at STAR by looking at forward di-pions in d+Au collisions.
Further insight into the uncorrelated pedestal below the near-side and away-side peaks in azimuthal correlations may be provided by differentiating between d+Au and p+Au collisions, by tagging on intact neutrons in the deuteron beam in d+Au collisions.
Comparisons to recent theories indicate that multi-parton interactions play a more significant role in d+Au collisions than p+Au collisions and offer a unique opportunity to study correlations between leading partons inside nucleons.
The general features found for the peaks in forward di-pion azimuthal correlations in d+Au collisions are also present in p+Au collisions.

\end{abstract}

\maketitle


\section{Introduction}

It is known that gluon densities in the proton rise for decreasing longitudinal partonic momentum fractions, ${\it x}$, however this rise cannot continue indefinitely for smaller ${\it x}$ values due to unitarity constraints.
Eventually, gluon recombination becomes important and non-linear contributions to evolution equations must be included, at which point the gluon density saturates \cite{MuellerQiu1986} \cite{McLerranVenugopalan1994}.
One model that describes these effects is the Color Glass Condensate (CGC) \cite{GribovLevinRyskin1983} \cite{MuellerQiu1986} \cite{McLerranVenugopalan1994} \cite{IancuVenugopalan2003} \cite{IancuLeonidovMcLerran2001} model which is a semi-classical, effective field theory used to describe low-${\it x}$ gluons within nuclei.
The gluon saturation regime can be reached for low ${\it x}$ values, large $\sqrt{s}$, large rapidities (y), and heavy nuclear targets (A).

By looking at forward rapidities at the Solenoidal Tracker at RHIC (STAR), asymmetric partonic collisions are probed, primarily involving high-${\it x}$ valence quarks in the probe colliding with low-${\it x}$ gluons in the target.
Therefore, the best opportunity to study saturation behavior at STAR is to look at forward rapidities using nuclear targets.
Perturbative QCD predicts that in standard 2-to-2 processes there is a high probability for back-to-back di-jets.
In a saturation picture, however, the $p_T$ of one jet can be balanced by multiple interactions with the dense gluon field in the target leading to a suppression of back-to-back jets \cite{KharzeevLevinMcLerran2005}.
Measurements of azimuthal correlations between two forward $\pi^0$ can probe a more limited, and smaller, ${\it x}$ range than inclusive measurements.

\section{Experimental Setup}
The STAR detector provides nearly hermetic coverage over a full 2$\pi$ azimuthal range and wide pseudo-rapidity range \cite{Braidot2011Thesis}.
The Beam-Beam Counter (BBC) was used to differentiate between central and peripheral d+Au collisions.
The Zero-Degree Calorimeter (ZDC) gives good neutron identification which can be used to tag intact neutrons from the deuteron beam.
The Barrel Electro-magnetic Calorimeter (BEMC) was used to identify neutral pions at mid-rapidity.
The Forward Meson Spectrometer (FMS) was used to trigger on neutral pions at forward-rapidities which is critical to reach the ${\it x}$ values needed to probe saturation effects.
The FMS is an array of lead glass cells with full azimuthal coverage providing electromagnetic calorimetry over a range of 2.5 < $\eta$ < 4.
The data in this analysis were collected during RHIC Run 8 at $\sqrt{s}$ = 200 GeV.

\section{Azimuthal Correlations in $p+p$ and $d+Au$ Collisions}

When triggering on a forward $\pi^0$, the rapidity of an associated $\pi^0$ is correlated with the $x_{bj}$ of the soft parton involved in the scattering.
Previously reported analyses \cite{Braidot2011Thesis} have measured azimuthal correlations between forward $\pi^0$ and both mid-rapidity $\pi^0$ and $h^\pm$ associated particles using the STAR BEMC which measure gluon densities at moderate ${\it x}$ ranges.
These analyses have shown no significant broadening in the away-side peak from p+p collisions to d+Au collisions and no hints of a disappearance of the away-side peak, indicating that this kinematic region does not reach the gluon saturation regime.

The lowest ${\it x}$ region can be reached at STAR by triggering on forward $\pi^0$ and measuring ${\it forward}$ associated $\pi^0$.
The previously reported analyses are not yet efficiency or background corrected but show a similar near-side peak in p+p and centrality averaged d+Au collisions \cite{Braidot2009QM}.
The away-side peak, however, is significantly broadened when comparing p+p collisions to d+Au collisions, hinting that the gluon saturation regime may have been reached.
Further evidence is provided by using the gold-side BBC charge sum to differentiate between central and peripheral d+Au collisions.
The away-side peak in peripheral d+Au collisions is similar to that seen in p+p collisions.
In central d+Au collisions, however, the away-side peak shows strong suppression and is in good agreement with CGC model calculations \cite{AlbaceteMarquet2010}.

\begin{figure}
  \includegraphics[width=0.32\textwidth]{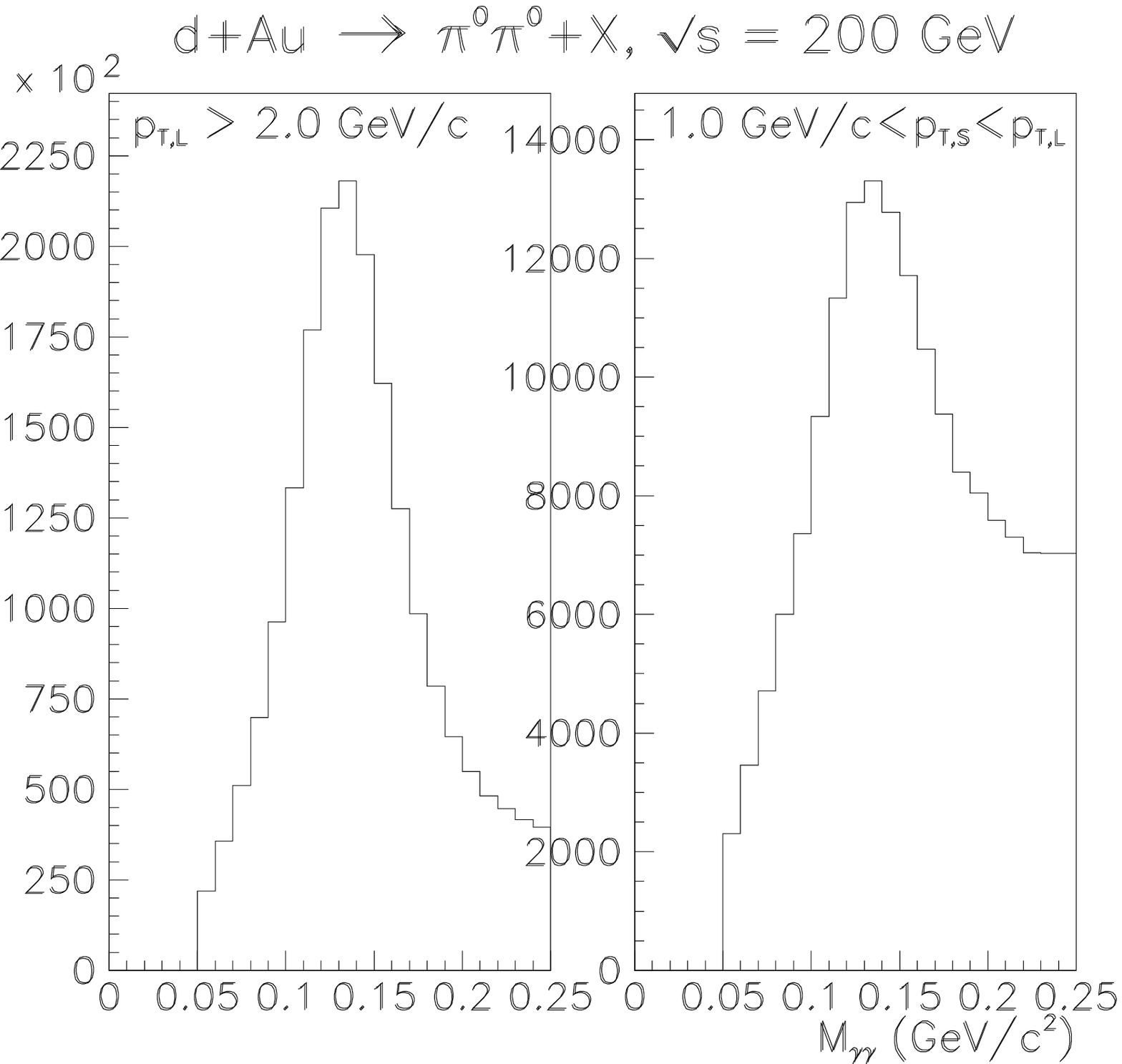}
  \includegraphics[width=0.32\textwidth]{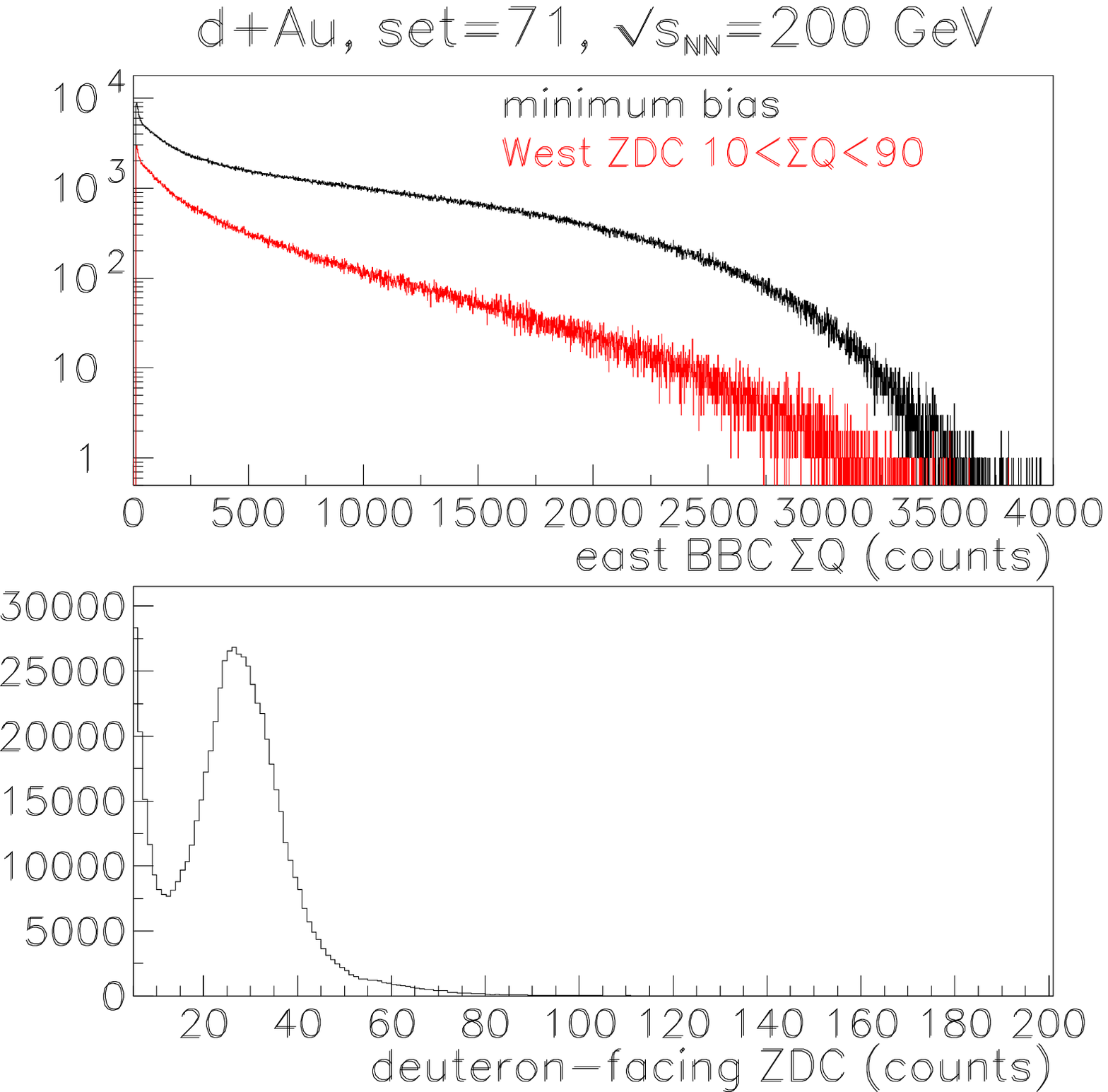}
  \includegraphics[width=0.32\textwidth]{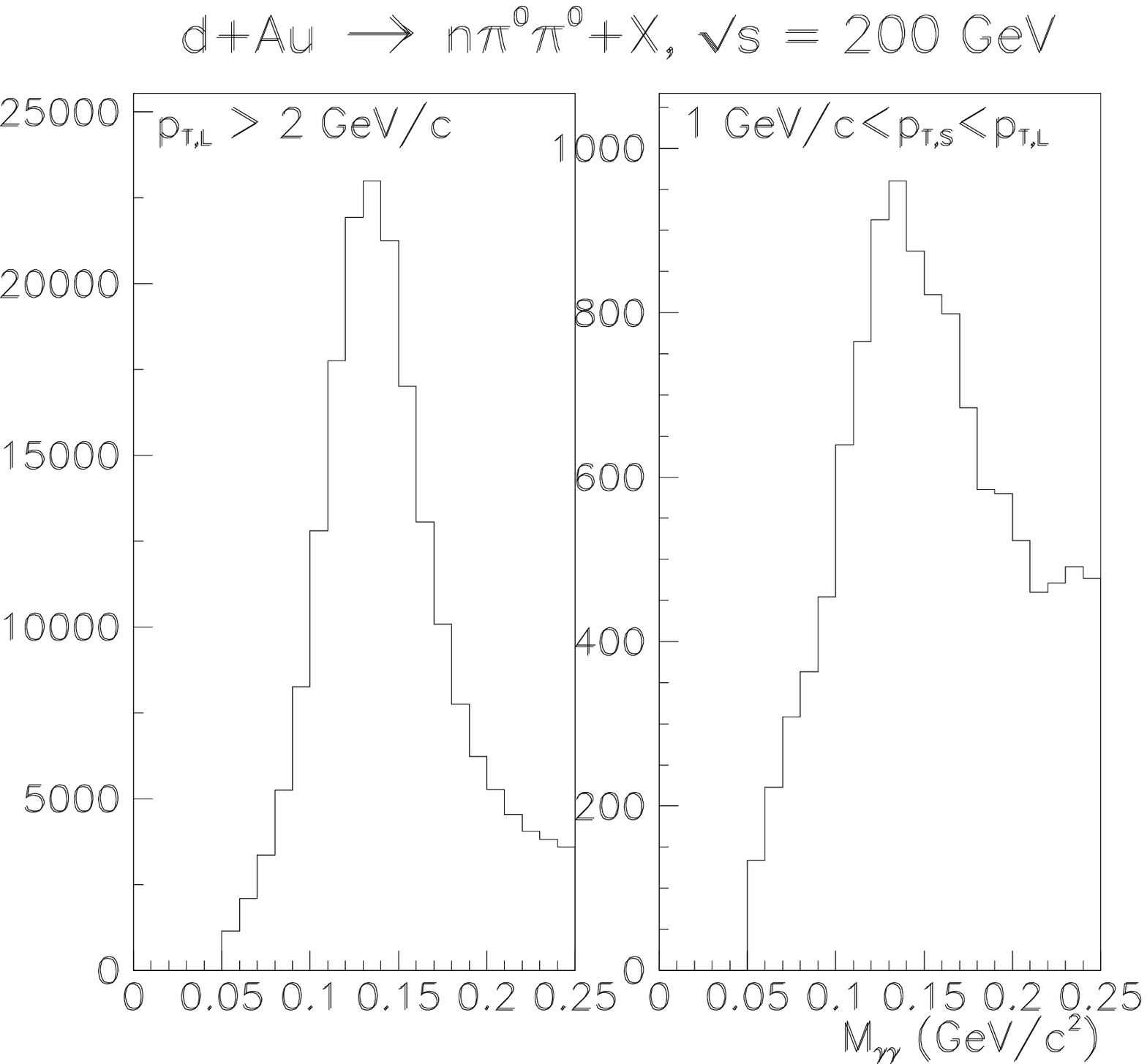}
  \caption{
(Left) Reconstructed $\pi^0$ invariant mass in d+Au collisions. Left plot shows leading $\pi^0$, Right plot shows sub-leading $\pi^0$.  
(Middle) Top plot shows gold-facing BBC charge sum distribution with and without neutron tagging for minimum bias events.  Bottom plot shows a clear spectator neutron peak for the deuteron-facing ZDC response. 
(Right) Reconstructed $\pi^0$ invariant mass in neutron tagged d+Au collisions. Left plot shows leading $\pi^0$, Right plot shows sub-leading $\pi^0$.}
  \label{mass_bbc_zdc}
\end{figure}

\begin{figure}
  \includegraphics[width=0.32\textwidth]{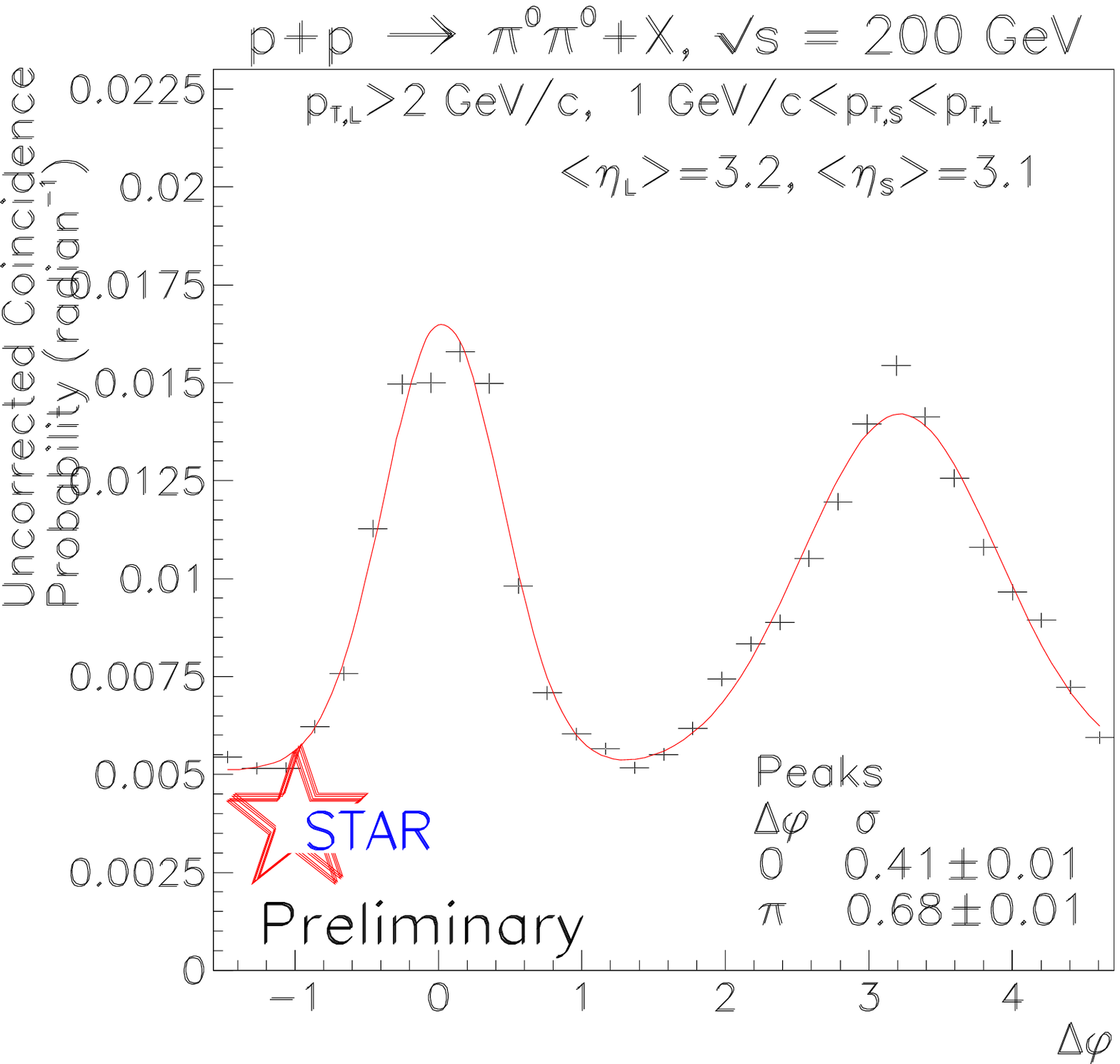}
  \includegraphics[width=0.32\textwidth]{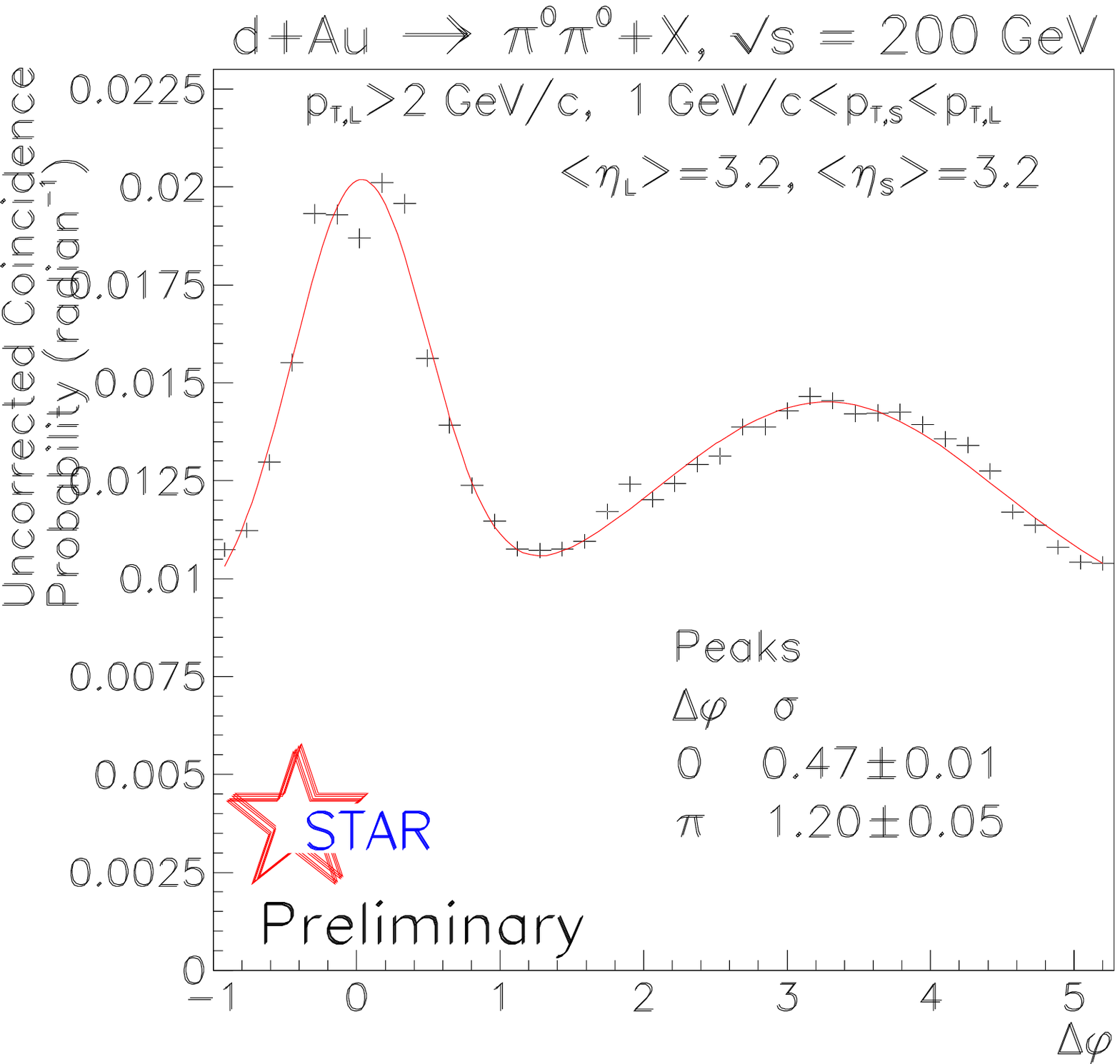}
  \includegraphics[width=0.32\textwidth]{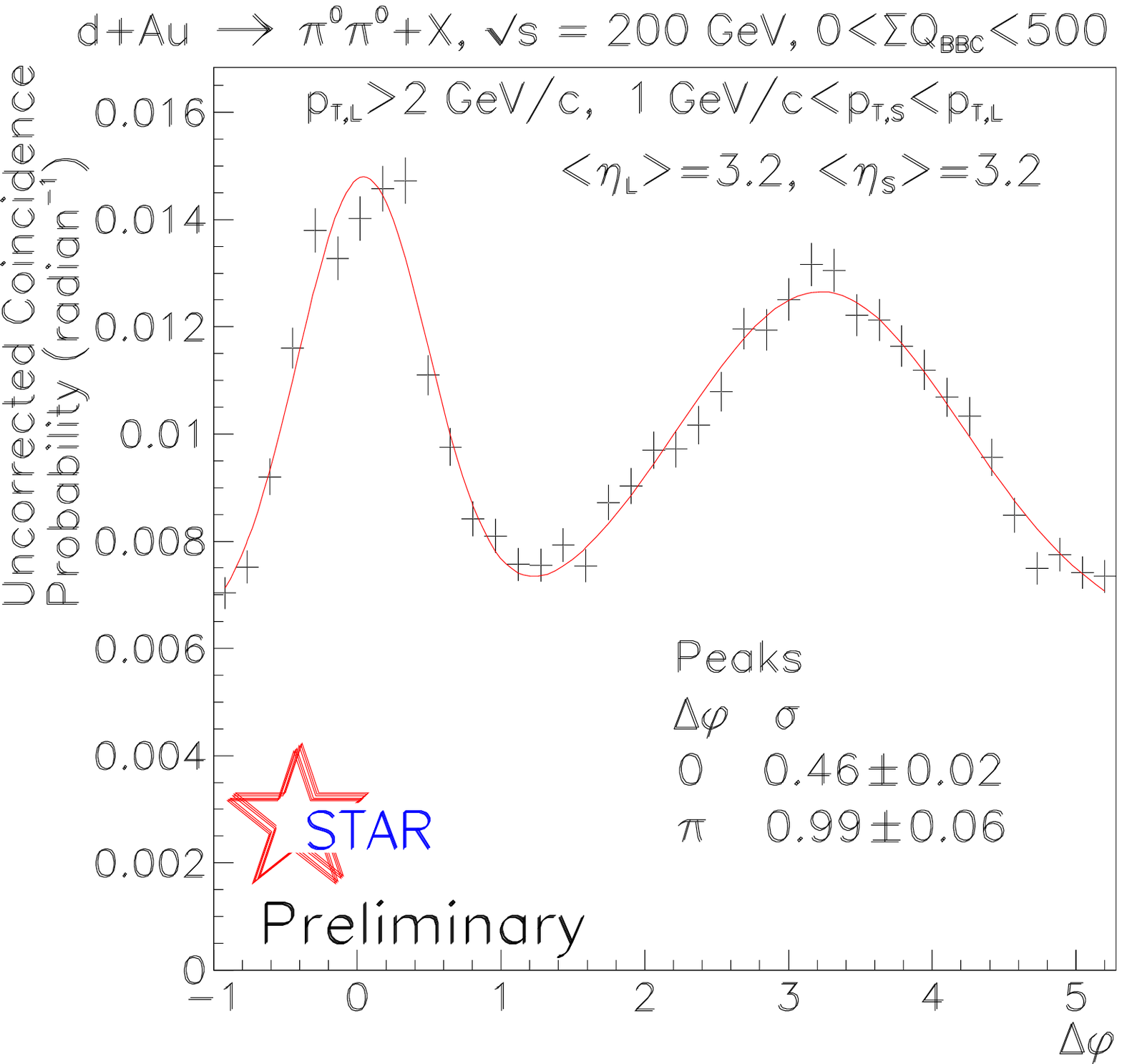}
  \caption{
Uncorrected coincidence probability versus azimuthal angle difference between two forward $\pi^0$ given a triggered forward $\pi^0$ at $\sqrt{s}$ = 200 GeV.
(Left) p+p collisions
(Middle) d+Au collisions, Centrality averaged
(Right) d+Au collisions, Peripheral events
}
  \label{pp_dau}
\end{figure}

\begin{figure}
  \includegraphics[width=0.39\textwidth]{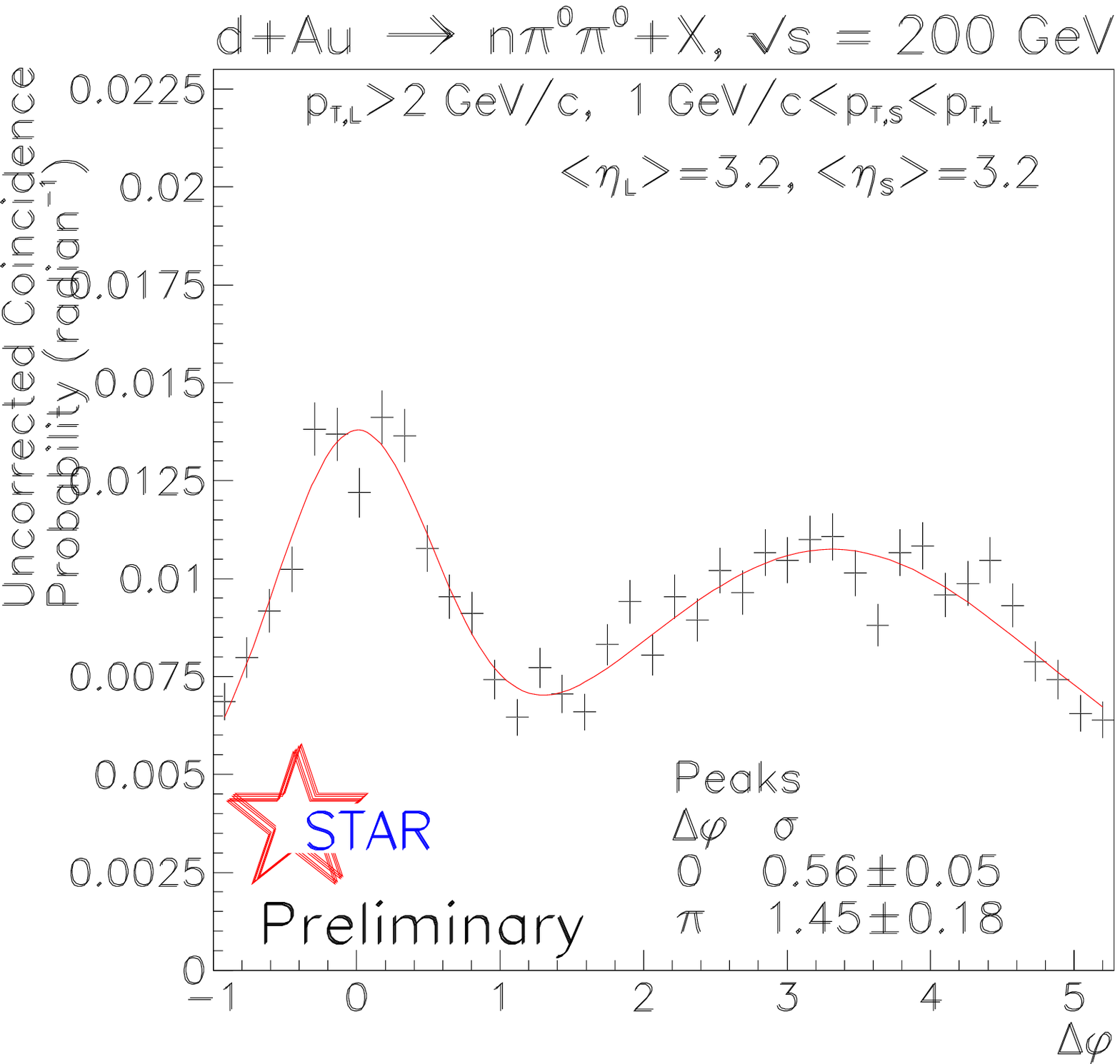}
  \includegraphics[width=0.39\textwidth]{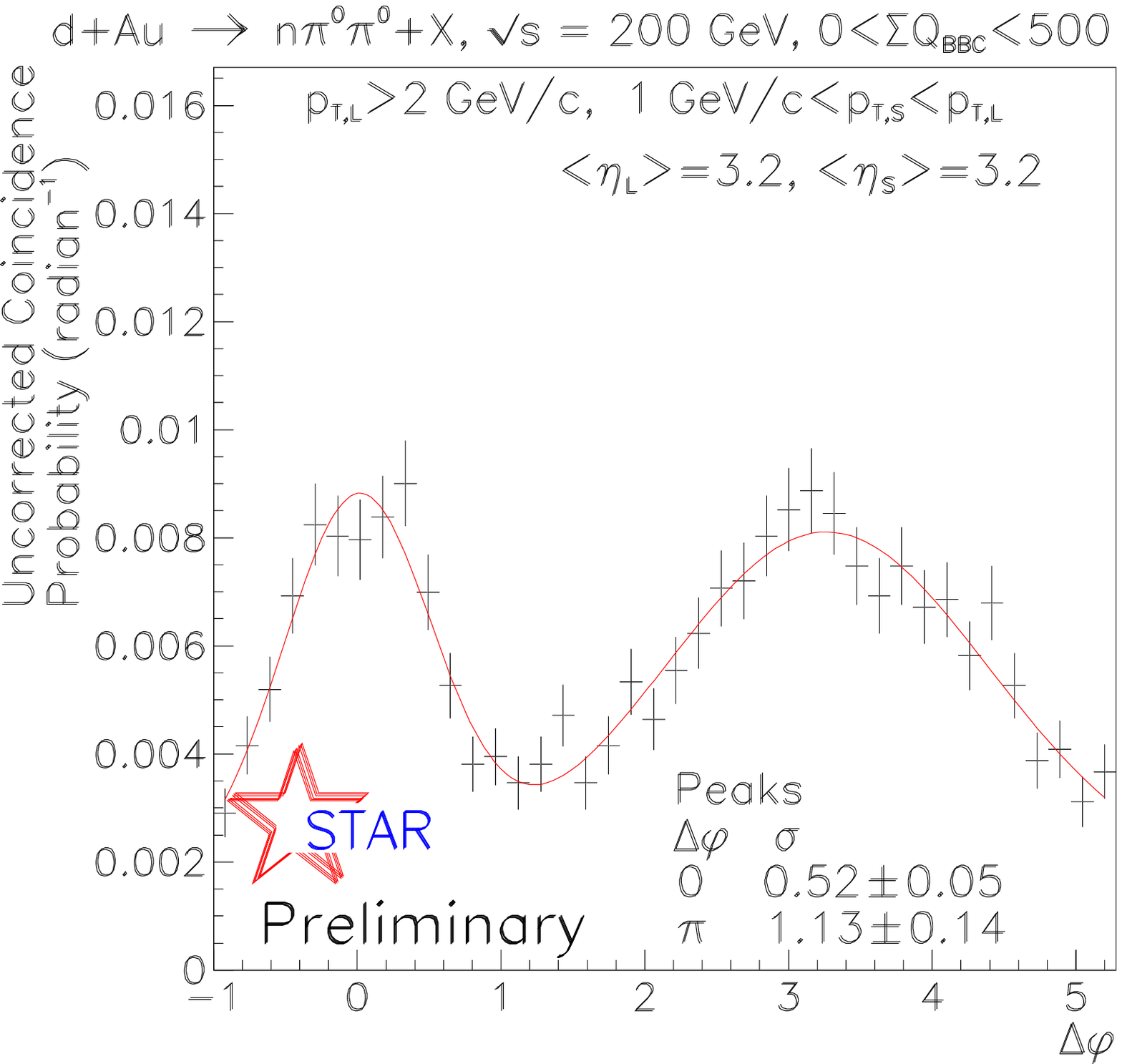}
  \caption{
Uncorrected coincidence probability versus azimuthal angle difference between two forward $\pi^0$ given a triggered forward $\pi^0$ in d+Au collisions with the described neutron tag (p+Au) at $\sqrt{s}$ = 200 Gev.
(Left) Centrality averaged
(Right) Peripheral events
}
  \label{pau}
\end{figure}

\section{Azimuthal Correlations in $p+Au$ Collisions}
It may also be useful to distinguish between p+Au and d+Au collisions by looking for events where the neutron in the deuteron beam stays intact.
A clear single-neutron peak can be seen in the deuteron-facing ZDC response for RHIC Run 8 d+Au Minimum Bias triggered data (See Figure~\ref{mass_bbc_zdc}) which can be used to select events where the neutron stays intact.
Cutting on this single-neutron peak introduces a slight bias towards peripheral collisions but a significant sample of central collisions remain.
The reconstructed di-pion invariant masses are shown in Figure~\ref{mass_bbc_zdc} with and without the neutron tag.
Because the invariant mass distributions look similar, it is expected that efficiency corrections of azimuthal correlations should be similar with and without neutron tagging although quantitative studies are still ongoing.

Uncorrected forward di-pion azimuthal correlations for centrality averaged and peripheral d+Au (no neutron tag) and p+Au (with neutron tag) collisions are shown in Figures~\ref{pp_dau} and~\ref{pau}.
The inclusion of the spectator neutron tag reduces the uncorrelated pedestal for both the centrality averaged and peripheral collisions.
The spectator neutron condition also has very little impact on the near-side and away-side peak heights above pedestal and peak widths.
A study of systematic errors is still in progress.

While most theories thus far have focused on peak heights and widths in di-pion azimuthal correlations, new theories have been put forth to explain the uncorrelated pedestal.
One theory postulates that the difference in pedestal level between p+p and d+Au collisions arrises due to multiple parton interactions in the d+Au collision \cite{StrikmanVogelsang2011}.
This theory includes both multiple scattering from one nucleon in the deuteron beam and one scattering from each nucleon on the deuteron beam.
The reduction of the uncorrelated pedestal from d+Au collisions to p+Au collisions indicates that multi-parton interactions play a more significant role in d+Au collisions.
Systematic study of the correlations between p+p, p+Au, and d+Au collisions are ongoing.
These observations could provide a unique window into correlations between leading partons within nucleons without having to use more complicated techniques such as double-scattering observables.

\section{Conclusions}
Previously reported analyses have shown no significant away-side peak broadening in forward-mid rapidity di-hadron azimuthal correlations.
Forward-forward rapidity di-pion azimuthal correlations have, however, shown significant broadening in the away-side peak between p+p and d+Au collisions and strong suppression of the away-side peak in central d+Au collisions.
Tagging intact spectator neutrons from the deuteron beam allows differentiation between p+Au and d+Au collisions which can give further insight into the uncorrelated pedestal.
Data appears to indicate that multi-parton interactions may contribute more to the pedestal in d+Au than p+Au collisions.
Other basic aspects of the azimuthal correlations are not significantly changed between d+Au and p+Au collisions.






\bibliographystyle{aipproc}   

\bibliography{sample}

\IfFileExists{\jobname.bbl}{}
 {\typeout{}
  \typeout{******************************************}
  \typeout{** Please run "bibtex \jobname" to optain}
  \typeout{** the bibliography and then re-run LaTeX}
  \typeout{** twice to fix the references!}
  \typeout{******************************************}
  \typeout{}
 }

\end{document}